\documentclass[prl,tighten,article,twocolumn,epsfig,graphics,showpacs,mathbbm]{revtex4}

\usepackage{amsmath,amsfonts,amssymb,graphics,graphicx,epsfig,color,times,bbm}

\begin{document}
\bibliographystyle{apsrev}

\newcommand{\R}{\mathbbm{R}}
\newcommand{\rr}{\mathbbm{R}}
\newcommand{\nn}{\mathbbm{N}}
\newcommand{\cc}{\mathbbm{C}}
\newcommand{\zz}{\mathbbm{Z}}

\newcommand{\ii}{\mathbbm{1}}

\newcommand{\id}{\mathbbm{1}}

\newcommand{\tr}[1]{{\rm tr}\left[#1\right]}
\newcommand{\gr}[1]{\boldsymbol{#1}}
\newcommand{\be}{\begin{equation}}
\newcommand{\ee}{\end{equation}}
\newcommand{\bea}{\begin{eqnarray}}
\newcommand{\eea}{\end{eqnarray}}
\newcommand{\St}{{\cal S}}
\newcommand{\Ad}{{\rm Ad}}

\newcommand{\ket}[1]{|#1\rangle}
\newcommand{\bra}[1]{\langle#1|}
\newcommand{\avr}[1]{\langle#1\rangle}
\newcommand{\G}{{\cal G}}
\newcommand{\eq}[1]{Eq.~(\ref{#1})}
\newcommand{\ineq}[1]{Ineq.~(\ref{#1})}
\newcommand{\sirsection}[1]{\section{\large \sf \textbf{#1}}}
\newcommand{\sirsubsection}[1]{\subsection{\normalsize \sf \textbf{#1}}}
\newcommand{\ack}{\subsection*{\normalsize \sf \textbf{Acknowledgements}}}
\newcommand{\front}[5]{\title{\sf \textbf{\Large #1}}
\author{#2 \vspace*{.4cm}\\
\footnotesize #3}
\date{\footnotesize \sf \begin{quote}
\hspace*{.2cm}#4 \end{quote} #5} \maketitle}
\newcommand{\eg}{\emph{e.g.}~}

\newcommand{\proofend}{\hfill\fbox\\\smallskip }

%---------------------------------------------------------------------------

\newtheorem{theorem}{Theorem}
\newtheorem{proposition}{Proposition}

\newtheorem{lemma}{Lemma}
\newtheorem{definition}{Definition}
\newtheorem{corollary}{Corollary}
\newtheorem{example}{Example}
\newtheorem{remark}{Remark}
\newtheorem{problem}{Problem}

\newcommand{\proof}[1]{{\it Proof.} #1 $\proofend$}

%\title{Computational difficulty of global variations
%in the density matrix renormalization group}

\title{Computational Difficulty of Global Variations in the Density Matrix Renormalization Group}

\author{J.\ Eisert}

\affiliation{
QOLS, Blackett Laboratory, 
Imperial College London,
London SW7 2BW, United Kingdom\\
Institute for Mathematical Sciences, 
Imperial College London,
London SW7 2PE, United Kingdom}

\date{\today}

\begin{abstract}
The density matrix renormalization group (DMRG)
approach is arguably the most successful method to numerically find ground states of quantum spin chains. It amounts to iteratively locally optimizing matrix-product states, aiming at better and better approximating the true ground state. To date, both a proof of convergence to the globally best approximation and an assessment of its complexity are lacking. Here we establish a result on the computational complexity of an approximation with matrix-product states: The surprising result is that when one globally optimizes over several sites of local Hamiltonians, avoiding local optima, one encounters in the worst case a computationally difficult NP-hard problem (hard even in approximation). The proof exploits a novel way of relating it to binary quadratic programming. We discuss intriguing ramifications on the difficulty of describing quantum many-body systems.
\end{abstract}

 
\maketitle

``How difficult is it to describe quantum systems in classical
terms''? This question in its various variants has
manifold implications to several fields of theoretical physics:
to the context of numerically
studying many-body systems of condensed-matter physics  
in their ground-state properties, to the question of the
superiority of a quantum compared 
to a classical computer, and others.
Recently, a renewed 
interest in questions of 
classically simulating quantum many-body systems 
[1--3] gave rise to a number of new results and  
simulation methods, a large number of them motivated 
or in their approach by
ideas of quantum information theory [2--8]. 
The arguable workhorse of numerically finding 
ground states of many-body systems, the  DMRG 
method [1--3], 
was recently reassessed and in 
some ways improved. It seems fair to say that the problem
of finding ground states 
of systems with periodic 
boundary conditions or 
higher-dimensional systems is now much better 
understood than not very 
long ago [6].
The performance of DMRG-type methods has also
been quantitatively related to  
entanglement scaling in ground states:  
One should expect an approximation in terms of
matrix-product states -- as DMRG is generating -- 
to be most faithful, if an ``entanglement area-theorem''
holds, which in turn is typically the case in non-critical systems 
[8, 10].

Now, even if matrix product states form the right set of
states that well-describe the true ground state properties,
and one can expect MPS to faithfully represent the ground state
\cite{Faithful}:
``How difficult is it then to find the truly best approximation to
the ground state''?
This is the key question
of the {\it computational complexity} of any method to find ground 
states. Quite surprisingly, given the maturity of the 
field and the significance of such simulations, 
this question is essentially open. In practice, DMRG 
produces very good results, despite of the possibility of 
getting stuck in local minima in the optimization. However,
it is not certifiable: one never can be entirely sure whether
one has indeed found a state close to the true ground state of the
system. Hence, to find certifiable methods to get ground states
seems very timely and important \cite{Osborne}.

\begin{figure}
\includegraphics[width=.45\textwidth]{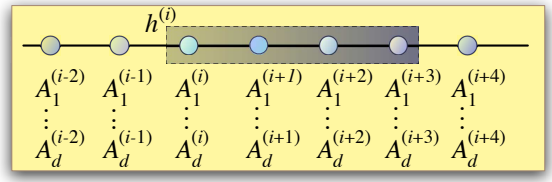}
\caption{Matrix product states and local hamiltonians.}\label{fig1}
\end{figure}

In this work, we present a first rigorous analysis of 
the complexity of finding ground states using variations
over matrix-product states as in DMRG. As such, 
DMRG essentially amounts to a local variation of matrices 
in the matrix-product states \cite{Scholl,Frank}. 
This is made most explicit in the variant of Ref.\
\cite{Frank}.
Here, we show that if 
we allow for a global variation over several sites at once -- 
to find the globally best approximation in this set and to
avoid local optima -- one encounters a problem which is
computationally hard, even in approximation. 
Or, actually more strongly: for any instance of a  binary
quadratic problem (including the NP hard 
exact satisfiability, or the maximum clique or independent
set  problems) 
one has an identical instance, realized with local 
translationally invariant hamiltonians \cite{Paris}.
To prove that this is 
true, we need $4$-level systems, $6$-local
hamiltonians, and a variation over two sites. 
This is much stronger than merely 
saying that polynomially constrained
problems of high degree as such are computationally
hard, as it could a priori of course 
well be that these hard instances never occur in the specific
context under consideration.
Moreover, this reduction can be found with polynomial
effort. So one faces the ironic and interesting situation
that when locally varying matrix product states, one has an
efficient subproblem, but may not get the certifiable
true ground state of the system. 
In turn, when aiming at
avoiding the problem of local minima and varying
matrices of several sites at once, one encounters a
hard problem.

{\it Local hamiltonians. --}
The considered hamiltonians are
$r$-local and translationally invariant, 
\begin{equation}\label{Ham}
	H= \sum_{i=1}^{n-r+1}  h^{(i)},
\end{equation}
up to open boundary conditions
(see Fig.\ 1), where the local interaction is governed by some
general hamiltonian
$h^{(i)} = \sum_{\alpha,\beta,\dots,\xi=1}^d
	h_{\alpha,\dots,\xi}
	(\sigma_\alpha^{(i)}\otimes 
	\sigma_\beta^{(i+1)}\otimes \dots\otimes
	\sigma_\xi^{(i+r-1)})$.
Here, $\{\sigma_\alpha^{(i)}\} $ denotes any 
local operator basis for site $i=1,\dots,n$, 
such as Pauli operators in case
of spin-$1/2$ systems. 
This insistence on local hamiltonians
renders the assessment of the computational complexity fair,
as in non-local models, one has the freedom
to incorporate frustrated higher-dimensional systems.
%for which one should not be surprised 
%when ground state approximation is computationally 
%difficult. 

{\it Matrix product states. --}
The density matrix renormalization group 
methods -- in  several variants -- essentially produce 
a sequence of 
matrix product states (MPS) that better and better approximate
the true ground state. These matrix product states
correspond to non-translationally invariant versions of 
the finitely correlated states \cite{FCS}, 
which were 
historically developed and studied independently from 
the DMRG context. 

We consider chains of $n$ sites and $d$-level constituents, 
so $d=2$ for a spin chain. The local basis is denoted by
$\{|1\rangle,\dots,|d\rangle\}$. Then, MPS take the
standard form 
%\begin{equation}\label{MPS}
	$|\psi\rangle =\sum_{i_1,\dots,i_n=1}^d
	A^{(1)}_{i_1} 
	A^{(2)}_{i_2} 
	\dots
	A^{(n)}_{i_n}  
	|i_1,i_2,\dots,i_n\rangle$.
%\end{equation}
Here,
$A^{(j)}_{i}\in \cc^{D_j\times D_{j+1}}$, 
$j=1,\dots,n$,   are complex matrices. 
These matrices, depending on the auxiliary 
dimension $D=\max_j D_j$ (the {\it MPS dimension}), 
characterize the MPS. 
For simplicity, we impose open boundary conditions.
This means that $D_1=D_{n+1}=1$, i.e.,
$A^{(n)}_1,\dots,A^{(n)}_d\in \cc^{D_n\times 1}$ 
and $A^{(1)}_1,\dots,A^{(1)}_d\in \cc^{1\times D_2}$
are taken to be vectors. 
We consider normalized MPS, meaning that 
$\langle \psi|\psi\rangle =1$.  This is notably
achieved by using the {\it gauge condition} that
%\begin{equation}\label{gauge}
	$\sum_{i} (A^{(j)}_{i})^\dagger  A^{(j)}_{i} =\id$
%\end{equation}
%(taken to be of rank $d$ for site $n$) 
for each site $j$ \cite{SiteN}. 
This condition can always be satisfied up to similarity
transformations \cite{Norm}, 
and is in numerical methods, most explicitly in
Ref.\ \cite{Frank}, insisted upon for numerical stability.

We look at optimizations over MPS, as it 
is explicitly or implicitly being done in 
DMRG approaches. 
Needless to say, there are many variants.
The local variational method for finite-size DMRG 
involves optimizations 
over single sites \cite{Frank}, essentially
equivalent to the $B\circ B$-method 
of DMRG in case of open boundary conditions:
For any site $j=1,\dots,n$, one keeps  
all matrices $A^{(i)}_{k}$
of the sites  $i$  different from $j$ fixed. Then, one 
finds the optimal MPS
by varying over the matrices $A^{(j)}_{1},\dots,
A^{(j)}_{d}$ of site $j$, satisfying normalization,
to minimize the energy
%\begin{equation}\nonumber
	$E=  \langle \psi | H | \psi\rangle$.
%\end{equation}
Then, one takes the next 
site, optimizes over the matrices of that
site,  and ``sweeps'', until a fixed point is 
reached \cite{Frank,Scholl,Finite}.
In fact, assertions that in gapped systems,
DMRG does find the ground state, implicitly assume
that the globally optimal MPS can be 
found. 

{\it Main result. --}
Yet, the problem one actually intends to solve is the 
full problem, so the global optimization
problem of finding the best matrix product state. Or, if one
has an infinite system, one should at least be able to 
solve the problem over several sites at once, in one run, 
solving the problem over these sites, say, of the length
scale of the classical correlation length. Obviously, we have
to allow for sequences of larger and larger systems in $n$,
as otherwise, the complexity question and the one of
finding the ground state no longer makes sense: for 
finite systems, there is always a finite $D$ 
to exactly write out the true ground state. 
We will now see that there are hard instances 
in the class of variations over several sites at once.
\begin{problem}[DMRG with global 
variation over several sites]
Consider a family of translationally invariant 
hamiltonians $H$ of the 
form as in Eq.\ (\ref{Ham}), with 
$n=a D +b $ ($a,b\in \rr$ are fixed numbers).
Let $I\subset \{1,\dots,n\}$ 
some finite subset of sites of the hamiltonian.
Then, find for any MPS dimension 
$D$ the optimal matrix product state
\begin{equation}\nonumber
	|\psi\rangle =\sum_{i_1,\dots,i_n=1}^d
	A^{(1)}_{i_1} 
	A^{(2)}_{i_2} 
	\dots
	A^{(n)}_{i_n}  
	|i_1,i_2,\dots,i_n\rangle
\end{equation}
by simultaneously 
varying the matrices $A^{(j)}_{1},\dots,
A^{(j)}_{d}$ of sites $j\in I$ 
satisfying the above gauge condition to
achieve normalization $\langle\psi|\psi\rangle=1$,
to minimize the energy
	$E=  \langle \psi | H | \psi\rangle$.
\end{problem}
This is the very reasonable and natural variation 
over several sites to minimize the energy,
aiming at avoiding local minima. Yet, surpringly,
one arrives at the subsequent observation:
\begin{theorem}[Hardness of 
DMRG with global variations] 
Finding the best MPS with varying over several sites
is NP-hard in $D$. Moreover, the ground state energy 
$\langle \psi | H|\psi\rangle$
can not be $1/D$-approximated 
in polynomial time.
\end{theorem}
The latter statement is meant unless P$=$NP 
\cite{Approximation,NP,Feige}. What we aim for is a reduction of 
this problem to a general binary quadratic problem. 
We start by abstract considerations, and then flesh out
how we can incorporate this in the MPS setting. 
To simplify the notation, we will first consider a setting 
involving $N$ real variables, $N$ even, and will later 
relate this to the  dimension $D$. 
What follows is not a standard
polynomial reduction to an NP-complete problem: we are
heavily restricted by the specific form offered by MPS. 
Yet, it will turn out that the freedom that we have is
just so sufficient for our purposes. We will introduce some 
new techniques.
Readers only interested in the physical implications may
read on the with the 
further discussion of simulatability issues.

{\it Idea of relating optimization problems. --}
We start by relating a certain class of quadratic 
continuous problems to a general binary quadratic problem, 
where variables can only take values in $\{0,1\}$,
which includes problems as the max clique problem.  This
will be the final form that we want to achieve with MPS, when
minimizing the energy $\langle\psi|H|\psi\rangle$.
The argument is as follows: For every real  $(N-1)\times
(N-1)$-matrix $M$ with entries having absolute values
smaller or equal to unity
there exists an $N\times N$-matrix $Y$ such that
every problem of finding the minimal value of
$ b M b^T$ for binary variables $b_1,\dots,b_{N-1} \in\{0,1\}$
can be written as a problem of finding the 
minimal value of
	$x Y y^T$
for $x_1,\dots,x_N,y_1,\dots,y_N\in[0,1]$.
%\end{lemma}
%
To see this, we may identify the first $N-1$ variables
$x_1,\dots,x_{N-1}$ with $b_1,\dots,b_{N-1}$. 
Now, for $x,y\in[0,1]^{N}$, 
\begin{eqnarray}\nonumber
	&& 2\sum_{k=1}^{N-1}
	x_k y_k -
	\sum_{k=1}^{N-1}
	(x_k + y_k ) -  x_N y_N\label{Big} \\
	&+& 
	(x_1 ,\dots,x_{N-1})
%	\left[
%	\begin{array}{cc}
	M 
%	& e^T\\
%	e & 0
%	\end{array}
%	\right]
	(x_1 ,\dots, x_{N-1} )^T/(2 (N-1)^2)\nonumber
%		=   -N ,\nonumber
\end{eqnarray}
%where $e=(1,\dots,1)/2$, 
takes its minimal value 
exactly if 
(i) $y_k=1-x_k$, $x,y\in\{0,1\}^N$,  (ii) $x_N=y_N=1$ hold,
and
(iii) $(x_1,\dots,x_{N-1})M (x_1,\dots,x_{N-1})^T$ takes its minimal
value over binary variables. 
Eq.\ (\ref{Big}) can clearly be incorporated in a single 
matrix $Y$ \cite{NiceIdea}.
In this form, we are in the position
to actually generate exactly this situation in the
MPS setting when minimizing the energy. Also, the above
approximation statement follows from results in Ref.\
\cite{Feige}, and using estimates for the deviation from binary
variables, once we are in the position of formulating
the problem as a minimization over $xY y^\dagger$.

{\it Incorporating this in MPS. --}
In order to generate a fair worst-case szenario, we are 
free to 
take any variational set $I$, fix the other matrices of sites
not contained in $I$ appropriately, and may 
take any local hamiltonian $H$ and local 
physical dimension. Then, energy minimization amounts
to solving the optimization problem. 
We will make use of a $6$-local hamiltonian, 
so $h$ acts non-trivially on $6$ subsystems, 
%\begin{equation}\label{actualham}
	$h^{(i)}=|1\rangle\langle 1|^{\otimes 6} 
	+ |2\rangle\langle 2|^{\otimes 6}
	+  |3\rangle\langle 3|^{\otimes 6}
	 + |4\rangle\langle 4|^{\otimes 6}$
%\end{equation}
for all $i$.
The idea now is to think of four ``regions'' of
the matrix product chain: To the left and to the right, 
there will be $m = \lceil \log_2 D\rceil$ sites, 
forming a ``tail'' to accumulate the proper range of 
the matrices. The left center consists of $6$ sites, 
and the system will be constructed in a way such that the
hamiltonian acts only non-trivially on these sites. The 
right center  is a chain
of sites generating ``indicator matrices''.
This means that the chain consists for integer $N$ 
of  $n=N^2+6+2m$ sites, and $D= 2 N^2+N$. 
$N$ labels both the auxiliary matrix
dimension and the system size (we can always 
pad the system to get $n= a D + b$ for $a,b\in \nn$). 

Let us first focus on the left center, embodying $6$ sites:
Sites $m+3$ and $m+5$ will form the set $I$, and we 
keep the other matrices fixed,
respecting the gauge condition. We
take 
%\begin{equation}\nonumber
	$A^{(m+4)}_1= 
	(\sum_{l=1}^N E(l,1) /N)
	 \oplus 0_{D-N}$,
%\end{equation}	 
and $A^{(m+4)}_2$ such that the gauge condition is satisfied,
the other two matrices of this site being zero.
Here, we use of the notation $E(i,j)$ for a matrix, 
all entries of which are zero, except
that $E(i,j)_{i,j}=1$.
This matrix $A^{(m+4)}_1$ has the
purpose of selecting appropriate 
parts of $A^{(m+3)}_1$ and $A^{(m+5)}_1$: the
matrices $A^{(m+3)}_1$ and $A^{(m+5)}_1$ will later
incorporate the variables $x_k$ and $y_k$, respectively.
It is not difficult to see that for any complex  $c_k, d_k$ satisfying 
$|c_k|\leq 1$ and $|d_k|\leq 1$, $k=1,\dots,N$,
we can find $D\times D$-matrices satisfying
$(A^{(m+3)}_1)^\dagger A^{(m+3)}_1 \leq \id_D$
and $(A^{(m+5)}_1)^\dagger A^{(m+5)}_1 \leq \id_D$, such that
\begin{equation}\label{rhs}
	A^{(m+3)}_1 A^{(m+4)}_1 A^{(m+5)}_1 =
	[
	c_1 , \dots , c_N
	]^\dagger\,
	[
	d_1 , \dots , d_N
	]/N.
\end{equation}
Conversely, for any solution of Eq.\ (\ref{rhs}) we find 
$|c_k |,|d_k | \leq1$ for all $k$.
This follows from exploiting the gauge conditions for 
$A^{(m+5)}_1$ and making use of the specific form 
of $A^{(m+3)}_1$ that has been chosen.
These numbers $c_1,\dots,c_N$ and $d_1,\dots,d_N$
are still complex: the key idea we make use of 
at this point is that we can appropriately combine them 
such that only
absolute values remain: the
role of the binary variables will be taken over
by 
	$x_k=|c_k|^2$, 
	$y_k=|d_k|^2$. 

{\it Generating indicator matrices from matrix products. --}
To the right of the left center, we will append the right center,
so appropriately chosen $N^2$ matrices.
As we insist on a local hamiltonian, we have no freedom to 
select only certain products: we will always have to deal
with all possible products. 
We hence have to
exploit a certain structure, such that from all 
exponentially many
products, we can generate polynomially many
indicator matrices, which are zero except 
from a {\it single non-zero element}.
These indicator matrices will be used in the MPS construction
to single out certain elements. It is not at all 
obvious that such matrices 
generating indicator matrices even exist:
Yet, for any $N\times N$ matrix $Y$ with entries
$Y_{k,l}\in [0,1)$, one can indeed 
construct $D\times D$-matrices 
$M_{1}^{(j)},M_2^{(j)}$ for $j=1,\dots , N^2$ 
with this property \cite{Construction}: 
For every binary word 
$(i_1,i_2,\dots,i_{N^2})\in \{0,1\}^{N^2}$,  we find
\begin{eqnarray}\nonumber
	P 
	\prod_{j=1}^{N^2} M_{i_j+1}^{(j)}
 	=\left\{
	\begin{array}{ll} 
	Y_{k,l}
	\left[
	\begin{array}{cc}
	E(k,l) & 0 \\
	0  & 0  \\
	\end{array}
	\right] & 
	\text{ or, }\\ 
	0, &  
	\end{array}\right.\label{R}
\end{eqnarray}
for $k,l=1,\dots,N$, where
$P\in \rr^{D\times D}$ is defined as
	$P=\sum_{k=1}^N\sum_{l=1}^N
	E(k+1,kN + l)$.
Also, each non-zero matrix in Eq.\ (\ref{R})
corresponds to a single binary word. 
In more colloquial and intuitive 
terms: we can take an arbitrary product of 
these matrices with lower index $1$ or $2$ defined by the
binary word, and multiply it from the left with $P$, acting
here as a shift operator. 
We will then always obtain a matrix with a single non-zero
element. This single element can be arbitrary,
forming the desired matrix $Y$. To check in retrospect 
that the given construction \cite{Construction} 
has this property is straightforward. 

{\it Combining results and minimizing the energy. --}
We are now in the position to put the previous results
together, and see that the binary optimization problem
can indeed be encoded in energy minimization. 
Concerning the further matrices of the left center, we simply
put $	A^{(m+1)}_1=A^{(m+6)}_1= \id_D$, and
$A^{(m+2)}_1=\id_N\oplus 0_{D-N}$, 
$A^{(m+2)}_2= 0_N\oplus \id_{D-N}$.
All other matrices of sites
$m+1,m+2,m+6$ are set to zero.
The matrices forming the right center
are identified with
	$A^{(m+6+l)}_{2+i}= M^{(l)}_i$,
	for 
	$l=1,\dots,N^2$,
	$i=1,2$,
and $A^{(m+6+l)}_{1}=A^{(m+6+l)}_{2}=0_{D}$ for all
$l=1,\dots,N^2$.  
The left and right tails are simply padded with matrices
such that the product of them gives rise to an
identity matrix \cite{Precise}.
Now, we find that we have satisfied the gauge
condition for all matrices of sites $\{1,\dots,n\}\backslash I$.
Finally, we can combine all this: 
Let the matrices associated with sites $\{1,\dots,n\}\backslash I$
be as chosen above, and the hamiltonian as in Eq.\ (\ref{Ham}). 
Then, energy minimization becomes
\begin{eqnarray}\nonumber
	\langle \psi | H |\psi\rangle =
	\sum_{k,l}
	|\text{tr}[A^{(m+3)}_1 A^{(m+4)}_1 A^{(m+5)}_1 Y_{k,l} 
	E(k,l)] |^2 = x Y y^T/N.
%	\nonumber \\
%	&=&  (x_1 ,\dots,x_{N})
%	Y (y_1 ,\dots, y_N )^\dagger.
\end{eqnarray}
This follows now from a direct evaluation of the overlap
of the MPS, making use of the above results.
This proves the validity of the theorem: the optimization problem
encountered in the variation over matrix-product state is in 
this form identical to solving the respective instance of the binary
quadratic problem, and for the very same hamiltonian, 
for every instance of this 
problem, as in an instance of 
the max clique problem, one can find a $D$ 
such that the variational problem becomes
identical. This shows that indeed: even within the setting of
varying over several sites simultaneously
when finding optimal matrix-product states to
approximate ground states, one encounters
computationally difficult NP hard problems.
 
{\it Further discussion of the simulatability of 
quantum many-body problems. --}
In this work, we have addressed the 
question of finding quantum ground states of
one-dimensional many-body systems on a classical
computer, as far as the complexity of local variations
is concerned. This is a question that has not explicitly
been addressed so far: even if
MPS are a set faithfully representing the true 
ground state, how difficult is it to find the optimal one?
It turns out that in the class of problems
where one reasonably 
varies over several sites at once
contains provably computationally difficult 
instances, even
for local one-dimensional hamiltonians.
By no means should this be read
as a statement that DMRG does not work: In practice, DMRG 
obviously gives typically rise to  very good results.
But rather as a warning sign, that to find best
approximations of many-body systems can be
computationally hard. Moreover, it suggests that to
further look for new certifiable algorithms to
find ground states, including ``error bars'',
at least for non-critical systems, 
should be a very fruitful  task. 
A number of questions
are implicitly raised here: What is the significance of 
breaking the translational symmetry in MPS
for ground state approximations?
What role does the gauge freedom
play? Also, what is the exact relationship to 
QMA completeness  \cite{Terhal}
of local hamiltonian problems? Can good
bounds be found via polynomial relaxations \cite{Relax}?
It would be exciting to further look at
truly optimal translationally invariant MPS as 
approximations of  translationally invariant ground 
states, 
to see under what conditions such approximations are truly
efficient. It is the hope that the present work further
fosters such considerations.

{\it Acknowledgements. --}
This work has benefited from discussions with 
F.\ Brand\~ao,
H.J.\ Briegel, 
H.A.\ Carteret,
J.I.\ Cirac,
M.\ Cramer,
W.\ D{\"u}r,
D.\ Gross,
A.\ Neumaier,
M.B.\ Plenio,
B.\ Terhal,
L.\ Tuncel, 
F.\ Verstraete,
R.F.\ Werner,
M.M.\ Wolf, and especially from
engaged and fun discussions with
T.J.\ Osborne. This work has been supported by the DFG
(SPP 1116, SPP 1078), the EU (QAP), 
the EPSRC, the QIP-IRC, 
Microsoft Research, and
the EURYI Award Scheme.


\begin{thebibliography}{99}
\bibitem{Osborne}
	Local properties can be found 
	in P for some non-critical systems  \cite{Osborne2}. For fixed
	system size and non-frustrated systems, quantum
	Monte Carlo techniques work certifiably \cite{MC}.
\bibitem{SiteN}
	To be very precise, 
	this is meant as the identity of appropriate rank, so 
	$\sum_{i=1}^d 
	(A_i^{(1)})^\dagger A_i^{(1)}= 
	\text{diag}(1,\dots,1,0,\dots,0)$ of rank $d$. As can
	readily be verified, successively exploiting these
	conditions, starting at site $n$, 
	any such MPS 
	is normalized \cite{Norm}.
\bibitem{NiceIdea}
	If one has $x,y \in[0,1]^{N-1} $ that satisfy
	$2\sum_{k=1}^{N-1}
	x_k y_k -
	\sum_{k=1}^{N-1}
	(x_k + y_k ) = -(N-1)$,
	then $x_k,y_k \in \{0,1\}$ for all 
	$k=1,\dots,N-1$.
	Hence, the binary character can be enforced via
	a single quadratic constraint: The first term ensures 
	that for each $k$ at least one of the variables
	$x_k$ and $y_k$ must vanish. The second 
	constraint is extremal, and can only be satisfied if
	for each $k$ either $x_k=1$ or $y_k=1$.
\bibitem{Approximation}
	For a function 
	$f:[0,1]^n\rightarrow \rr$ to be minimized 
	denote by $f^*(n)$ and $f_*(n)$ the 
	maximum and the minimum
	of $f$. An algorithm is then a 
	$\mu$-approximation, if it computes 
	$\tilde f(n)$ satisfying $\tilde f (n)- f_\ast (n)
	\leq \mu(n) (f^* (n)- f_*(n))$ \cite{NP}.
\bibitem{Construction}
	Very explicitly, the construction is as follows: Let
	$M^{(N (k-1) +l)}_1 = \left[
		\begin{array}{cc}
		0_{N} &   0_{N,D-N}\\
		G_{k,l} & 0_{N,D-N}\\
		H_{k,l} & 0_{N,D-N} 
		\end{array}
	\right]$,
%	\begin{equation}\nonumber
%		M^{(N (k-1) +l)}_1 = \left[
%		\begin{array}{cc}
%		0_{N} &   0_{N,D-N}\\
%		G_{k,l} & 0_{N,D-N}\\
%		H_{k,l} & 0_{N,D-N} 
%		\end{array}
%	\right],
%\end{equation}
for $k,l=1,\dots,N$, where 
	$G_{k,l}= Y_{k,l} 
	E(k(N-1)+l,l)\in \rr^{N^2\times N}$, and
	$M^{(N (k-1) +l)}_2=  0_{N} \oplus \id_{D-N}$.
	The matrices $H_{k,l}\in \rr^{N^2\times N}$
	can be chosen such that 
	$(M_1^{(i)})^\dagger M_1^{(i)}
	+
	 (M_2^{(i)})^\dagger M_2^{(i)}
	 =\id_{D}$ for all $i$.
\bibitem{Proof}
	The product $M_1^{(i)} M_1^{(j)}$ 	
	vanishes
	whenever $i\neq j$. Hence, the only contributing
	binary words are of the form $(1,\dots,1,0,1,\dots,1)$ with a 
	single zero element, and $(1,\dots,1)$.
\bibitem{Precise}
	Take
	$A_3^{(k)}= \sum_{l=1}^{2^{k-1}}E(l,l)$
	and $ A_4^{(k)}= \sum_{l=1}^{2^{k-1}}E(2^{k-1}+l,l)$
	for $k=1,\dots,m$, the other matrices of sites
	$1,\dots,m$ being zero. The same matrices are taken in 
	reverse order for $n-m,\dots,n$.
\end{thebibliography}
\end{document}